\documentclass[10pt,a4paper]{article}
\usepackage{graphicx}
\usepackage{amsmath}
\usepackage{amssymb}
\usepackage{float}

\begin{document}

\title{\textbf{A Model Explaining Correlation Between Observed Values in Contingency Tables}}
\author{\\Abhik Ghosh, Samit Roy, Sujatro Chaklader \\
B. Stat. 3rd Year, \\ Indian Statistical Institute}
\date{}
\maketitle

\begin{abstract}
In this article, a model is proposed using Bayesian techniques to account for the high correlation between many observed set of contingency tables. In many real life data this high correlation is encountered. Simulation studies are also given to check the effectiveness of this model.
\end{abstract}
\tableofcontents
\clearpage
\section{Introduction}
Consider the following contingency table-
\begin{table}[h]
	\centering
		\begin{tabular}{|c|c|c|c|}\hline
			  & Success & Failure & Total \\ \hline
			trt 1 & $y_1$ & $n_1-y_1$ & $n_1$ \\ \hline
			trt 2 & $y_2$ & $n_2-y_2$ & $n_2$ \\ \hline  
		\end{tabular}
		\caption{contingency table}
\end{table}
\bigskip
\\
For this type of data, we do not know the individual responses for the treatments but only the aggregate values. But from a set of such tables we can find out the correlation between the observations. In some cases, the observed correlation was as high as $.7$ or $.8$. For example, consider the following data~-
\begin{table}[h]
	\centering
		\begin{tabular}{c c c c c}\hline
			Trial&\multicolumn{2}{c}{Treatment}&\multicolumn{2}{c}{Control} \\ \hline
			$i$&$x_{i}^{T}$&$n_{i}^{T}$&$x_{i}^{C}$&$n_{i}^{C}$ \\ \hline
			1& 2& 39& 1& 43 \\ 
			2& 4& 44& 4& 44 \\
			3& 6& 107& 4& 110 \\
			4& 7& 103& 5& 100 \\
			5& 7& 110& 3& 106 \\ \hline
		\end{tabular}
		\caption{\emph{Outcome data('T' for treatment group and 'C' for control group) for prophylactic use of Lidocaine after heart attack (AMI)
(Hine et al. (1989), following Normand (1999))}}
\end{table}
\bigskip
\\ For these data, the correlation between treatment and control group turns out to be $0.9565$.
\\ But in the usual method for analysis of this kind of data and calculating the odds ratio etc., the underlying model assumed for the individual observations are \emph{independent} bernoullie with a common success probability. So, the calculations become easier and the aggregate values follow a binomial distribution with same success probability. But, although easier to calculate, this model cannot account for the high correlation values encountered that was mentioned earlier. For this problem, other models for explaining this type of data must be ventured.
 At first, we considered a marcov chain type model, i.e. for the individual observations, we considered that the observations depends only on the previous observation, but not on any other observations. But though this gives some correlation between the individual values, it is not useful for analysis of real life data. Because, the assumption that a obsrvation depends only on the previous observation is not encountered in practice and thus this model was not good for practical application. \\
Then, we tried a Bayesian approach. We introduced an error random variable in the success probabilities of the individuals and tried to explain the high correlation. This method was quite good for actual data analysis. The model is desribed in the following section
\section{The Model}
The general trend for analysis of this kind of data is to assume the individual observations to be independent and recalling the notations of table 1, $y_1$ follows $Bin(n_1,p_1)$ and $y_2$ follows $Bin(n_2,p_2)$ independently where $p_1$ and $p_2$ are the individual success probabilities for trt 1 an trt 2 respectively. But, as this cannot explain the correlation, we take an additional latent variable $\delta$ and try to rewrite the model that can explain the correlation\\
\subsection{The Model Containing Latent Variable}
 Consider $\delta$ as an latent variable associated with each observation of a contingency table---a simple choice for the distribution of $\delta$ can be taken as $N(0,1)$. Then assume that given $\delta$, individual observations have success probabilities, a function of $\delta$, $p_1(\delta)$ for trt 1 and $p_2(\delta)$ for trt 2 where 
$$p_1(\delta)=\frac{e^{\alpha_1+\delta}}{1+e^{\alpha_1+\delta}} \mathrm{\ and \ } p_2(\delta)=\frac{e^{\alpha_2+\delta}}{1+e^{\alpha_2+\delta}}$$ \\
Here $\alpha_1$ and $\alpha_2$ are two parameters corresponding to the success probabilities $p_1$ and $p_2$ of trt 1 and trt2 respectively. So, 
$$\alpha_1=\alpha_2 \Leftrightarrow p_1(\delta)=p_2(\delta)$$Then given $\delta$, $y_1$ follows $Bin(n_1,p_1(\delta))$ and $y_2$ follows $Bin(n_2,p_2(\delta))$ independently. But, due to introduction of the same latent variable $\delta$ for both the cases, their unconditional distributions are not independent. We have verified that this model can explain high as well as low correlation between $\frac{y_1}{n_1}$ and $\frac{y_2}{n_2}$ [Section 3.1]. In this model, the true success probabilities $\pi_1$ and $\pi_2$ of the treatments $1$ and $2$ can be given by
\begin{eqnarray}
\pi_i=E_{\delta}[p_i(\delta)]=\int{\frac{e^{\alpha_i+\delta}}{1+e^{\alpha_i+\delta}}}f(\delta|y_1,y_2)d\delta ~~~~~ \mathrm{for }~i=1,2.
\end{eqnarray}

\subsection{Estimation of Parameters}
So, the joint likelihood for $\alpha_1$, $\alpha_2$ and $\sigma^{2}$ is given by
\begin{flushleft}
\begin{eqnarray}
&L(\alpha_1,\alpha_2,\sigma^{2})=f(y_1,y_2,\delta | \alpha_1,\alpha_2,\sigma^{2})\nonumber \\
&=(\stackrel{n_1}{y_1})p_1(\delta)^{y_1}(1-p_1(\delta))^{n_1-y_1}(\stackrel{n_2}{y_2})p_2(\delta)^{y_2}(1-p_2(\delta))^{n_2-y_2}\frac{1}{\sqrt{2\pi \sigma}}e^{-\frac{\delta^{2}}{2\sigma^{2}}}
\end{eqnarray} 
\end{flushleft} 

To find the maximum likelihood estimates of $\alpha_1$, $\alpha_1$ and $\sigma^{2}$, we maximize equation $(2)$. The MLEs obtained are shown below
\begin{eqnarray*}
\frac{e^{\hat{\alpha_1}+ \delta}}{1+e^{\hat{\alpha_1}+ \delta}}=\frac{y_1}{n_1} &\Rightarrow& \hat{\alpha_1}=log\left( \frac{y_1}{n_1-y_1}\right)-\delta \\
\mathrm{and\ } \frac{e^{\hat{\alpha_2}+ \delta}}{1+e^{\hat{\alpha_2}+ \delta}}=\frac{y_2}{n_2} &\Rightarrow& \hat{\alpha_2}=log\left( \frac{y_2}{n_2-y_2}\right)-\delta \\
\mathrm{and\ } \hat{\sigma^{2}}=\delta^{2}
\end{eqnarray*}
But $\delta$ is not explicitly observed here. So, we use the Expectation-Maximization(EM) Algorithm to find $\hat{\alpha_1}$, $\hat{\alpha_2}$ and $\hat{\sigma^{2}}$. As in EM Algorithm, we replace $\delta$ by its expected value i.e., 
$$E(\delta|y_1,y_2)=\int{\delta f(\delta|y_1,y_2)d\delta}$$
where the conditional density of $\delta$ given $y_1$ and $y_2$ is given by
$$f(\delta|y_1,y_2)\propto(\stackrel{n_1}{y_1})\frac{e^{y_1(\alpha_1+\delta)}}{(1+e^{(\alpha_1+\delta)})^{n_1}}(\stackrel{n_2}{y_2})\frac{e^{y_2(\alpha_2+\delta)}}{(1+e^{(\alpha_2+\delta)})^{n_2}}\frac{1}{\sqrt{2\pi \sigma}}e^{-\frac{\delta^{2}}{2\sigma^{2}}}$$
Because of such a complicated form, the calculation of expected value of $\delta$ becomes very difficult. So, we replace $\delta$ by numerically calculated average value of $\delta$ given $y_1$ and $y_2$. This is known as \emph{Generalised EM Algorithm}. More specifically, we replace $\delta$ by 
$$\frac{1}{M}\sum_{i=1}^{M}{\delta_i}$$
where $\delta_i$'s are i.i.d. observations simulated from conditional density of $\delta$ given $y_1$ and $y_2$ i.e., from $f(\delta|y_1,y_2)$. Due to the complex nature of this density function, we opt for Metropolis-Hastings algorithm to generate the $\delta_i$'s. M is the number of $\delta_i$'s generated and it should be quite large for obtaining good results.Based on a suitable convergence criteria, we estimate $\hat{\alpha_1}$, $\hat{\alpha_2}$ and $\hat{\sigma^{2}}$. \\
But our goal is to estimate the success probabilities $\hat{\pi_1}$ and $\hat{\pi_2}$. We can estimate these two values using equation $(1)$ and the estimates $\hat{\alpha_1}$, $\hat{\alpha_2}$ and $\hat{\sigma^{2}}$ as follows~-
\begin{eqnarray}
\hat{\pi_1}&=&\int{\frac{e^{\hat{\alpha_1}+\delta}}{1+e^{\hat{\alpha_1}+\delta}}}f(\delta|y_1,y_2)d\delta \\
\mathrm{and\ } ~~\hat{\pi_2}&=&\int{\frac{e^{\hat{\alpha_2}+\delta}}{1+e^{\hat{\alpha_2}+\delta}}}f(\delta|y_1,y_2)d\delta 
\end{eqnarray}
Equations $(3)$ and $(4)$ are quite complex and exact calculation is not possible. So, the estimates of $\hat{\pi_1}$ and $\hat{\pi_2}$ are calculated numerically again using Metropolis-Hastings algorithm.\\
Now, to find the variance of the estimators $\hat{\alpha_1}$ and $\hat{\alpha_2}$, we compute the information matrix corresponding to the parameters $\alpha_1$ and $\alpha_2$ as follows ~-
$$I(\alpha_1,\alpha_2) = \begin{pmatrix}
-n_1p_1(\delta)(1-p_1(\delta)) & 0 \\ 0 & -n_2p_2(\delta)(1-p_2(\delta)) \end{pmatrix}$$
Note that, the only random part in the information matrix comes from $\delta$. So, taking expectation of $-I(\alpha_1,\alpha_2)$ with respect to the distribution of $\delta$ we get the dispersion matrix of $\hat{\alpha_1}$ and $\hat{\alpha_2}$. The covariance between $\hat{\alpha_1}$ and $\hat{\alpha_2}$ turns out to be $0$ and the variances are given by
\begin{eqnarray}
Var(\hat{\alpha_i})&=&E_{\delta}[n_ip_i(\delta)(1-p_i(\delta))]\\ \nonumber
&=&\int{\frac{n_1e^{\alpha_i+\delta}}{(1+e^{\alpha_i+\delta})^2}}f(\delta|y_1,y_2)d\delta ~~~~~ \mathrm{for }~i=1,2.
\label{va}
\end{eqnarray}
We can estimate these variances using above equations replacing $\alpha_i$ by $\hat{\alpha_i}$ for $i=1,2$ and $\sigma$ by $\hat{\sigma}$ and then using Metropolis-Hastings probabilities to evaluate the integrals.

\subsection{Odds Ratio}
The odds ratio is a measure of effect size, describing the strength of association or non-independence between two binary data values. It is used as a descriptive statistic, and plays an important role in logistic regression. Unlike other measures of association for paired binary data such as the relative risk, the odds ratio treats the two variables being compared symmetrically, and can be estimated using some types of non-random samples.
Now, when $\delta$ is known the Odds Ratio(OR) is given by
\begin{eqnarray*}
\theta&=&\frac{p_1(\delta)(1-p_2(\delta))}{p_2(\delta)(1-p_1(\delta))}\\
\Rightarrow log(\theta)&=&log\left( \frac{p_1(\delta)}{1-p_1(\delta)}\right)-log\left( \frac{p_2(\delta)}{1-p_2(\delta)}\right)\\
&=&(\alpha_1+\delta)-(\alpha_2+\delta)\\
&=&\alpha_1-\alpha_2
\end{eqnarray*}
Note that the value of $log(\theta)$ does not depend on the value of $\delta$ so that we can use the above formula  to compute the log odds ratio in our model where $\delta$ is actually unknown.Thus a estimator of the log odds ratio $log(\theta)$ can be 
$$\hat{log(\theta)}=\hat{\alpha_1}-\hat{\alpha_2}$$
Here, the association between the two treatments depends only on $\alpha_1$ and $\alpha_2$ --- the two parameters determining the succsess probabilities of the treatments and the independence of the two treatments is equivalent to the condition 
$$log(\theta)=\alpha_1-\alpha_2=0 \mathrm{\ or \ } \alpha_1=\alpha_2$$
Also note that the variance of this estimator is given by
\begin{eqnarray*}
Var(\hat{log(\theta)})&=&Var(\hat{\alpha_1}-\hat{\alpha_2})\\
&=&Var(\hat{\alpha_1})+Var(\hat{\alpha_2})\\
&=&E_{\delta}[n_1p_1(\delta)(1-p_1(\delta))]+E_{\delta}[n_2p_2(\delta)(1-p_2(\delta))]
\end{eqnarray*}
So we can estimate $Var(\hat{log(\theta)})$ as
$$\widehat{Var(\hat{log(\theta)})}=\widehat{Var(\hat{\alpha_1})}+\widehat{Var(\hat{\alpha_2})}$$
where $\widehat{Var(\hat{\alpha_1})}$ and $\widehat{Var(\hat{\alpha_2})}$ are obtained from previous section.

\subsection{Test of Independence}
As, mentioned in the previous section, testing for independence is equivalent to testing
$\alpha_1=\alpha_2$ or $log(\theta)=0$ \\
So the null hypothesis of independence $H_0:\alpha_1=\alpha_2$ against any suitable alternative can be tested by the Test Statistic ~-
$$T=\frac{\hat{log(\theta)}}{\sqrt{\widehat{Var}(\hat{log(\theta)})}}=\frac{\hat{\alpha_1}-\hat{\alpha_2}}{\sqrt{\widehat{Var}(\hat{\alpha_1}-\hat{\alpha_2})}}$$
Note that , the exact distribution of $T$ is not known but since $\hat{\alpha_1}$, $\hat{\alpha_2}$ and are MLE, the asymptotic distribution of $T$ is normal with mean $\alpha_1$-$\alpha_2$ and variance $1$.Thus under $H_0$, $T$ follows $N(0,1)$ and we can perform the test against any alternative.For example,we reject $H_0$ against $H_1 : log(\theta)\neq 0$ at level $\alpha$ if the observed value of $|T|$ is bigger than the upper ${\frac{\alpha}{2}}^{th}$ quantile of standard normal distribution.

\subsection{Generalization for k Tables}
In the previous sections, we only discussed a single table. But how to modify the calculation when k tables are given from k different study with the same treatments? There may be several ways for such modifications, the simplest one being using a weighted average of the estimates obtained from each table.So, at first we obtain $\hat{\alpha_{1i}}$, $\hat{\alpha_{2i}}$, $\hat{\pi_{1i}}$, $\hat{\pi_{2i}}$ and $\hat{\sigma_i}^{2}$ from the $i^{th}$ table as before for $i=1,2,...k$. Then, we use these values to get better estimates as~-
$$\hat{\alpha_{1}}^{(k)}=\frac{\sum_{i=1}^{k}{n_{1i}\hat{\alpha_{1i}}}}{\sum_{i=1}^{k}{n_{1i}}},\hat{\alpha_{2}}^{(k)}=\frac{\sum_{i=1}^{k}{n_{2i}\hat{\alpha_{2i}}}}{\sum_{i=1}^{k}{n_{2i}}} \mathrm{\ and \  }\hat{\sigma}^{2(k)}=\frac{\sum_{i=1}^{k}{n_i\hat{\sigma_{i}}^{2}}}{\sum_{i=1}^{k}{n_i}}$$
Here the superscript $(k)$ indicates the estimators obtained using $k$ tables.Similarly, $\hat{\pi_j}^{(k)}$ are weighted average of $\hat{\pi_{ji}}$'s over $i$ for $j=1,2$. \\
The variances of $\hat{\alpha_j}^{(k)}$ for $j=1,2$ are then given by 
\begin{eqnarray*}
Var(\hat{\alpha_j}^{(k)})&=&Var\left(\frac{\sum_{i=1}^{k}{n_{ji}\hat{\alpha_{ji}}}}{\sum_{i=1}^{k}{n_{ji}}}\right)\\
&=&\frac{\sum_{i=1}^{k}{n_{ji}^2Var(\hat{\alpha_{ji}})}}{(\sum_{i=1}^{k}{n_{ji}})^2}\\
&=&\frac{\sum_{i=1}^{k}{n_{ji}^2E_{\delta}[n_{ji}p_j(\delta)(1-p_j(\delta))]}}{(\sum_{i=1}^{k}{n_{ji}})^2}
\end{eqnarray*}

For this case the modified estimate of log of OR will be 
\begin{eqnarray*}
(\hat{log\theta})^{(k)}&=&\hat{\alpha_1}^{(k)}-\hat{\alpha_2}^{(k)}\\
&=&\frac{\sum_{i=1}^{k}{n_{1i}\hat{\alpha_{1i}}}}{\sum_{i=1}^{k}{n_{1i}}}-\frac{\sum_{i=1}^{k}{n_{2i}\hat{\alpha_{2i}}}}{\sum_{i=1}^{k}{n_{2i}}}
\end{eqnarray*}
with 
\begin{eqnarray*}
Var(\hat{log(\theta)}^{(k)})=\frac{\sum_{i=1}^{k}{n_{1i}^2E_{\delta}[n_{1i}p_1(\delta)(1-p_1(\delta))]}}{(\sum_{i=1}^{k}{n_{1i}})^2}+\frac{\sum_{i=1}^{k}{n_{2i}^2E_{\delta}[n_{2i}p_2(\delta)(1-p_2(\delta))]}}{(\sum_{i=1}^{k}{n_{2i}})^2}
\end{eqnarray*}
These variances of $\hat{\alpha_j}^{(k)}$'s and $\hat{log(\theta)}^{(k)}$ can be estimated using Metropolis-Hastings Algorithm as in the section 2.2.
\section{Simulation Study}
\subsection{Explaining Correlation}
We have done various simulations to see the effectiveness of our proposed model in explaining the correlation between $\frac{y_1}{n_1}$ and $\frac{y_2}{n_2}$ . For this purpose, we have simulated $k$ number of tables with a fixed value of $\alpha_1$ and $\alpha_2$ for each table, and possibly different values of $n_1$ and $n_2$. We simulated $10,000$ such sets of tables from our proposed model and computed the correlation between $\frac{y_1}{n_1}$ and $\frac{y_2}{n_2}$ for each sets of tables.The histogram plots of two such sets of correlations are shown in figure (1) and figure (2).
\begin{figure}[H]
\begin{center}
\includegraphics[width=8cm]{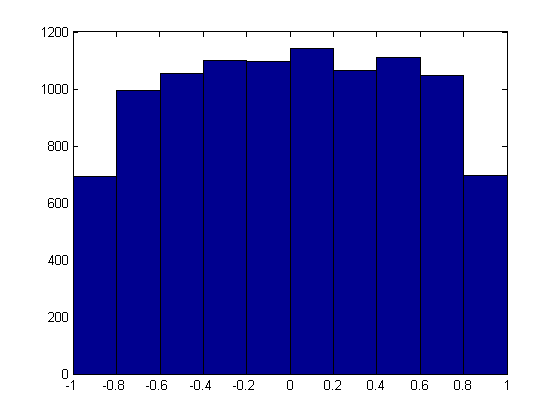}
\caption{Histogram plot of the correlation obtained from simulated data with proposed model(Study 1)}
\end{center}
\end{figure}
\begin{figure}[H]
\begin{center}
\includegraphics[width=8cm]{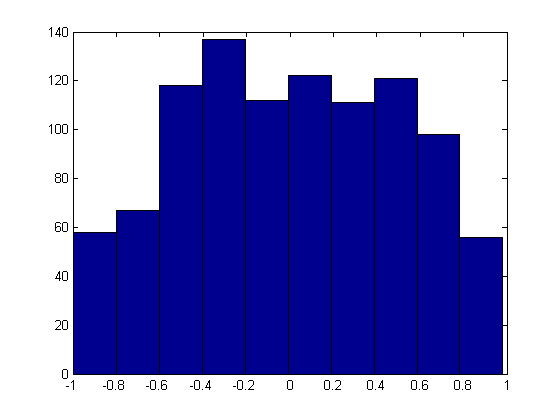}
\caption{Histogram plot of the correlation obtained from simulated data with proposed model(Study 2)}
\end{center}
\end{figure}
From these figures we can see that the $95\%$ quantile value of the distribution of the correlation is  around 0.8.Tables with high correlations have a significant probability under our proposed model and so this model can explain the high correlation within its significance region.Hence the practical data-sets of contingency tables having high correlation can be explained efficiently by the given model.

\subsection{Performance of the Proposed Method}
To observe the performance of the prposed method of estimation and testing, we have simulated $k$ number of tables with a fixed value of $\alpha_1$ and $\alpha_2$ for each table, and possibly different values of $n_1$ and $n_2$. Then, we have estimated $\hat{\alpha_1}$,$\hat{\alpha_2}$,$\hat{\pi_1}$, $\hat{\pi_2}$, $\hat{log(\theta)}$ and their variances by the method discussed in the previous sections for each of the tables separately and also using the
set of all the $k$ tables. We have also tested for independence first for each tables separately and then using the complete set of $k$ tables. We have done this for a large number of times for different values of $\alpha_1$ and $\alpha_2$. Some of the results that we have obtained are shown in table(3). 
\begin{table}[h]
	\centering
		\begin{tabular}{|c|c|c|c|c|c|c|c|}\hline
			$Trial$&$\hat{\alpha_1}$&$\hat{\alpha_2}$&$\hat{\pi_1}$&$\hat{\pi_2}$&$\hat{log(\theta)}$&$Test$&$Test$ \\ 
			&(s.e.) & (s.e.)& & &(s.e.) & Stat.~T& Result\\ \hline \hline
1	&	-0.2231	&	-0.7156	&	0.4454	&	0.3301	&	0.4925	&	0.1976	&	0	\\
	&	(2.2132)	&	(1.1474)	&		&		&	(2.4929)	&		&		\\\hline
2	&	0.033	&	-0.1676	&	0.5069	&	0.4573	&	0.2006	&	0.0738	&	0	\\
	&	(1.574)	&	(2.2181)	&		&		&	(2.7198)	&		&		\\\hline
3	&	0.3768	&	-1.9005	&	0.5972	&	0.1337	&	2.2773	&	0.7636	&	0	\\
	&	(2.6764)	&	(1.3155)	&		&		&	(2.9822)	&		&		\\\hline
4	&	0.3896	&	-0.5913	&	0.5967	&	0.3593	&	0.9809	&	0.3738	&	0	\\
	&	(1.091)	&	(2.3866)	&		&		&	(2.6241)	&		&		\\\hline
5	&	-0.7007	&	0.8397	&	0.3346	&	0.6988	&	-1.5404	&	0.6624	&	0	\\
	&	(1.8212)	&	(1.4461)	&		&		&	(2.3255)	&		&		\\\hline \hline
	
All&	-0.0174	&	-0.5597	&	0.4987	&	0.3829	&	0.5423	&	0.3401	&	0	\\
	&	(1.2138)	&	(1.0339)	&		&		&	(1.5945)	&		&		\\ \hline

		\end{tabular}
		\caption{\emph{Results of the simulation study with $\alpha_1=0$, $\alpha_2=0$, $\sigma=1$ and $n1=[20~ 10~ 30~ 5 ~15]'$, $n2=[6~ 20~ 15~ 25~10]'$ (The test result is $1$ if the null hypothesis of independence is rejected and $0$ otherwise)}}
\end{table}

Note that for the above table the data are simulated with a proposed model with $\alpha_1=0$ and $\alpha_2=0$ so that the expected success probabilities $\pi_1$ and $\pi_2$ are 0.5 each.As we can see from the table that the estimated success probabilities turns out to be very close to 0.5 and the test based on $T$ is always accepted implying the true structure of the equality of $\alpha_i$'s.Also the standard error of the estimates are less when using the combined data with all the tables compared to that obtained from individual tables.Similar results holds for other simulated tables also.This shows that our proposed method performs quite better for the contingency tables with possible explanation for its high correlation.
\newpage
\section{Analysis of the Real Deata}
Finally we apply our proposed method to analyse two sets of real data :
\begin{enumerate}
	\item Outcome data('T' for treatment group and 'C' for control group) for prophylactic use of Lidocaine after heart attack (AMI)
(Hine et al. (1989), following Normand (1999)) given in table(2)
	\item Outcome data for treatment group of a multicenter clinical trial (with high
sparsity) (Cancer and Leukemia Group, Cooper et al. (1993)) given in table(4)
\end{enumerate}
For these two sets of tables, the correlation between $\frac{y_1}{n_1}$ and $\frac{y_1}{n_1}$ are observed to be $0.9565$ and   $0.0562$ i.e., first one gives high correlation whereas the second one gives very low correlation. We have analysed these two tables and have seen that for high as well as low correlation, our proposed method gives reasonably better results. For both sets of tables, we have computed $\hat{\alpha_1}$,$\hat{\alpha_2}$,$\hat{\pi_1}$, $\hat{\pi_2}$, $\hat{log(\theta)}$ and their variances by the method discussed in the previous sections for each of the tables separately and also using the
set of all the tables. We have also tested for independence first for each tables separately and then using the complete set of all tables. The results obtained are shown in table$(5)$ and table$(6)$.
\begin{table}[ht]
	\centering
		\begin{tabular}{c c c c c}\hline
			Trial&\multicolumn{2}{c}{Treatment}&\multicolumn{2}{c}{Control} \\ \hline
			$i$&$x_{i}^{T}$&$n_{i}^{T}$&$x_{i}^{C}$&$n_{i}^{C}$ \\ \hline
1	&	1 &3& 3& 4	\\
2	&	8 &11& 3 &4	\\
3	&	2 &3& 2& 2	\\
4	&	2 &2& 2& 2	\\
5	&	0& 3& 2& 2	\\
6	&	2 &3& 1& 3	\\
7	&	2 &3 &2 &2	\\
8	&	4 &4& 1& 5	\\
9	&	2 &3 &2 &2	\\
10	&	2& 3& 0& 2	\\
11	&	3 &3 &3 &3	\\
12	&	0& 2& 2& 2	\\
13	&	1 &5 &1 &4	\\
14	&	2 &4& 2& 3	\\
15	&	4 &6 &2 &4	\\
16	&	3 &9 &4& 12	\\
17	&	2 &3 &1 &2	\\
18	&	1 &4& 3 &3	\\
19	&	2 &3 &1 &4	\\
20	&	0& 2 &0& 3	\\
21	&	1 &5 &2 &4	\\
 \\ \hline
		\end{tabular}
		\caption{\emph{Outcome data for treatment group of a multicenter clinical trial (with high
sparsity) (Cancer and Leukemia Group, Cooper et al. (1993))}}
\end{table}

\begin{table}[h]
	\centering
		\begin{tabular}{|c|c|c|c|c|c|c|c|}\hline
			$Trial$&$\hat{\alpha_1}(s.e.)$&$\hat{\alpha_2}(s.e.)$&$\hat{\pi_1}$&$\hat{\pi_2}$&$\hat{log(\theta)}(s.e.)$&$Test$&$Test$ \\ 
			& & & & & & Stat.~T& Result\\ \hline \hline
1	&	-2.9205	&	-3.7404	&	0.0512	&	0.0232	&	0.8199	&	0.4844	&	0	\\
	&	(1.3752)	&	(0.9866)	&		&		&	(1.6925)	&		&		\\ \hline 
2	&	-2.2995	&	-2.2995	&	0.0912	&	0.0912	&	0	&	0	&	0	\\
	&	(1.909)	&	(1.909)	&		&		&	(2.6997)	&		&		\\ \hline
3	&	-2.8231	&	-3.2769	&	0.0561	&	0.0364	&	0.4538	&	0.1471	&	0	\\
	&	(2.3795)	&	(1.963)	&		&		&	(3.0847)	&		&		\\ \hline
4	&	-2.6189	&	-2.9449	&	0.068	&	0.05	&	0.326	&	0.0971	&	0	\\
	&	(2.5543)	&	(2.1794)	&		&		&	(3.3577)	&		&		\\ \hline
5	&	-2.6944	&	-3.5417	&	0.0633	&	0.0281	&	0.8473	&	0.2762	&	0	\\
	&	(2.5528)	&	(1.7021)	&		&		&	(3.0682)	&		&		\\ \hline \hline
Combined	&	-2.688	&	-3.2069	&	0.0644	&	0.0422	&	0.0422	&	0.3493	&	0	\\
	&	(1.1713)	&	(0.9135)	&		&		&	(1.4854)	&		&		\\ \hline

		\end{tabular}
		\caption{\emph{Results obtained from the first data set(The test result is $1$ if the null hypothesis of independence is rejected and $0$ otherwise)}}
\end{table}

The table (5) shows the results for the data $1$ for prophylactic use of Lidocaine after heart attack (AMI) having treatment group and control group.The results obtained shows that the probability of success for the treatment group is around 0.0644 which is quite close to the success probabilities around 0.0422 of control group.So, as expected, the test based on any table accept the null hypothesis of the equality of the success probabilities for the treatment and control groups indicating that the prophylactic use of Lidocaine after heart attack (AMI)have no significant effect for the given treatment.

The table (6) shows the results for the data $2$ for treatment group of a multicenter clinical trial (with high
sparsity.The results obtained shows that the two success probabilities are around 0.525 and 0.5385.Also the test of independence or the equality of the success probabilities for the two groups is rejected for $8$ tables whereas it is accepted for the rest $13$ table which ,on an average,should indicate the acceptance of the null and this is obtained using the data on all the $21$ tables.
\\

\begin{table}[h]
	\centering
		\begin{tabular}{|c|c|c|c|c|c|c|c|}\hline
			$Trial$&$\hat{\alpha_1}(s.e.)$&$\hat{\alpha_2}(s.e.)$&$\hat{\pi_1}$&$\hat{\pi_2}$&$\hat{log(\theta)}(s.e.)$&$Test$&$Test$ \\ 
			& & & & & & Stat.~T& Result\\ \hline \hline
1	&	-0.6944	&	1.0973	&	0.3331	&	0.7498	&	-1.7917	&	1.5052	&	0	\\
	&	(0.8163)	&	(0.8663)	&		&		&	(1.1903)	&		&		\\ \hline
2	&	0.9822	&	1.1	&	0.7277	&	0.7504	&	-0.1178	&	0.0688	&	0	\\
	&	(1.4763)	&	(0.8655)	&		&		&	(1.7113)	&		&		\\ \hline
3	&	0.6899	&	2.9924	&	0.6648	&	0.9519	&	-2.3025	&	2.6343	&	1	\\
	&	(0.8172)	&	(0.31)	&		&		&	(0.874)	&		&		\\ \hline
4	&	2.9939	&	2.9939	&	0.9522	&	0.9522	&	0	&	0	&	0	\\
	&	(0.3091)	&	(0.3091)	&		&		&	(0.4371)	&		&		\\ \hline
5	&	-3.4002	&	2.9968	&	0.0323	&	0.9524	&	-6.397	&	14.5984	&	1	\\
	&	(0.3112)	&	(0.3085)	&		&		&	(0.4382)	&		&		\\ \hline
6	&	0.6988	&	-0.6875	&	0.6684	&	0.3352	&	1.3863	&	1.2007	&	0	\\
	&	(0.8153)	&	(0.8175)	&		&		&	(1.1545)	&		&		\\ \hline
7	&	0.6936	&	2.9962	&	0.6668	&	0.9524	&	-2.3026	&	2.6382	&	1	\\
	&	(0.8164)	&	(0.3085)	&		&		&	(0.8728)	&		&		\\ \hline
8	&	3.6766	&	-1.3986	&	0.9753	&	0.1982	&	5.0752	&	5.3704	&	1	\\
	&	(0.3142)	&	(0.8912)	&		&		&	(0.945)	&		&		\\ \hline
9	&	0.693	&	2.9956	&	0.6666	&	0.9524	&	-2.3026	&	2.6378	&	1	\\
	&	(0.8165)	&	(0.3086)	&		&		&	(0.8729)	&		&		\\ \hline
10	&	0.693	&	-2.9446	&	0.6666	&	0.05	&	3.6376	&	4.168	&	1	\\
	&	(0.8165)	&	(0.3082)	&		&		&	(0.8727)	&		&		\\ \hline
11	&	3.3941	&	3.3941	&	0.9675	&	0.9675	&	0	&	0	&	0	\\
	&	(0.3122)	&	(0.3122)	&		&		&	(0.4415)	&		&		\\ \hline
12	&	-2.9456	&	2.9945	&	0.05	&	0.9523	&	-5.9401	&	13.6183	&	1	\\
	&	(0.3081)	&	(0.3087)	&		&		&	(0.4362)	&		&		\\ \hline
13	&	-1.3827	&	-1.095	&	0.2006	&	0.2507	&	-0.2877	&	0.2309	&	0	\\
	&	(0.8954)	&	(0.8668)	&		&		&	(1.2463)	&		&		\\ \hline
14	&	0	&	0.6931	&	0.5	&	0.6667	&	-0.6931	&	0.5369	&	0	\\
	&	(1)	&	(0.8165)	&		&		&	(1.291)	&		&		\\ \hline
15	&	0.6931	&	0	&	0.6667	&	0.5	&	0.6931	&	0.4537	&	0	\\
	&	(1.1547)	&	(1)	&		&		&	(1.5275)	&		&		\\ \hline
16	&	-0.6924	&	-0.6924	&	0.3335	&	0.3335	&	0	&	0	&	0	\\
	&	(1.4144)	&	(1.6332)	&		&		&	(2.1605)	&		&		\\ \hline
17	&	0.6931	&	0	&	0.6667	&	0.5	&	0.6931	&	0.6417	&	0	\\
	&	(0.8165)	&	(0.7071)	&		&		&	(1.0801)	&		&		\\ \hline
18	&	-1.0988	&	3.401	&	0.2501	&	0.9677	&	-4.4998	&	4.8898	&	1	\\
	&	(0.8661)	&	(0.3111)	&		&		&	(0.9203)	&		&		\\ \hline
19	&	0.6947	&	-1.0971	&	0.6671	&	0.2503	&	1.7918	&	1.5053	&	0	\\
	&	(0.8162)	&	(0.8664)	&		&		&	(1.1903)	&		&		\\ \hline
20	&	-2.9469	&	-3.3698	&	0.05	&	0.0333	&	0.4229	&	0.966	&	0	\\
	&	(0.3082)	&	(0.3109)	&		&		&	(0.4378)	&		&		\\ \hline
21	&	-1.3863	&	0	&	0.2	&	0.5	&	-1.3863	&	1.0333	&	0	\\
	&	(0.8944)	&	(1)	&		&		&	(1.3416)	&		&		\\ \hline \hline
Combined	&	0.1485	&	0.3783	&	0.525	&	0.5385	&	-0.2298	&	0.5438	&	0	\\
	&	(0.2889)	&	(0.3083)	&		&		&	(0.4225)	&		&		\\ \hline

		\end{tabular}
		\caption{\emph{Results obtained from the second data set(The test result is $1$ if the null hypothesis of independence is rejected and $0$ otherwise)}}
\end{table}

\section{Discussion}
As we have seen in our discussion that in real life data, the correlation between the treatment effects are sometimes considerably high. The general independent binomial model cannot account for the high correlation. We tried to propose a model to account for these correlations. For this purpose we introduced a latent random variable associated with each observation giving rise to the correlation. For this model, we describe the an efficient procedure to estimate the parameters and the variances in estimation and also formulated a test for independence based on $log$ of Odds Ratio. Simulation studies showed that the correlationm obtained in our proposed model has a distribution with a large variance and $95\%$ quantile being around $.8$. So, cases with very high correlations have a significant probability under our proposed model and thus can be explained by this model. Simulation studies also show that our estimation procedure gives reasonably good estimates with low variance.\\

Note that, the test discussed here is based on large sample approximation of the null distribution of the test statistics and hence it may not be much reliable for small sample sizes. But the given test statistics can be used for small sample sizes also if we use the exact null distribution of the test statistics which can be obtained by a simulation study.\\
Concluding our discussion we would like to mention that using different distributions for the latent variable, we may come up with a different conclusion which may be better or worse than our model. So, further study on this topic can be done to observe how the change in the distribution of the latent variable affects the model and the estimation procedure. Also, work can be done to obtain the exact distribution of the test statistics discussed here to find an efficient test for small sample sizes.

\section{Acknowledgement}
We would like to thank Prof. Atanu Biswas and Prof. Bimal Roy for their valuable guidance throughout this project.

\end{document}